\documentclass[useAMS,usenatbib]{mn2e}
\usepackage{times}
\usepackage{graphics,epsfig}
\usepackage{graphicx}
\usepackage{amssymb}
\usepackage{txfonts}

\newcommand{\kms}{km\,s$^{-1}$} 

\newcommand{\nhi}  {\ensuremath{N}_{\rm HI}}

\newcommand{\cm}{cm$^{-2}$}
\newcommand{\hi}{H{\sc i}}

\newcommand{\hii}{H{\sc i}\,21cm}
\newcommand{\mhi}{M$_{\rm HI}$}

\title[The gas mass of low-$z$ DLAs]{Constraints on the gas masses of low-$z$ damped Lyman-$\alpha$ systems}
\author[Mazumdar et al.]{Parichay Mazumdar$^{1,2}$,
Nissim Kanekar$^{2}$\thanks{E-mail: nkanekar@ncra.tifr.res.in} and J. Xavier Prochaska$^{3}$\\
$^{1}$St. Stephen's College, University Enclave, New Delhi 110007, India\\
$^{2}$National Centre for Radio Astrophysics, Tata Institute of Fundamental Research, Ganeshkhind, Pune 411007, India\\
$^{3}$UCO/Lick Observatory, UC Santa Cruz, Santa Cruz, CA 95064, USA}

\begin{document}
\date{Accepted yyyy month dd. Received yyyy month dd; in original form yyyy month dd}


\maketitle

\label{firstpage}

\begin{abstract}
We report a deep search for redshifted H{\sc i}\,21cm emission from three damped and sub-damped 
Lyman-$\alpha$ absorbers (DLAs) at $z \approx 0.1$ with the Green Bank Telescope (GBT). 
No evidence for a redshifted H{\sc i}\,21cm emission signal was obtained in the GBT spectra of 
two absorbers, with the data on the third rendered unusable by terrestrial interference. The 
non-detections of H{\sc i}\,21cm emission yield strong constraints on the H{\sc i} masses of 
the associated galaxies, M$_{\rm HI} < 2.3 \times 10^9 \times (\Delta V/100)^{1/2}$~M$_\odot$ 
for the sub-DLA at $z = 0.0830$ towards J1553+3548, and M$_{\rm HI} < 2.7 \times 10^9 \times 
(\Delta V/100)^{1/2}$~M$_\odot$ for the DLA at $z = 0.0963$ towards J1619+3342, where 
$\Delta V$ is the H{\sc i}\,21cm line width, in km~s$^{-1}$. This continues the trend of low H{\sc i} 
masses found in all low-$z$ DLAs and sub-DLAs that have been searched for redshifted H{\sc i}\,21cm 
emission. Low-redshift absorbers with relatively low H{\sc i} column densities, 
$\lesssim few \times 10^{20}$~cm$^{-2}$, thus do not typically arise in massive gas-rich galaxies.

\end{abstract}


\section{Introduction}
\label{sec:intro}

Understanding the nature of the galaxies that give rise to damped Lyman-$\alpha$ systems (DLAs) 
and their redshift evolution has been an important question in galaxy evolution for nearly 
thirty years. With \hi\ column densities $\geq 2 \times 10^{20}$~\cm, 
similar to those seen on sightlines through local galactic disks, DLAs are the highest \hi\ 
column density systems detected in absorption in the optical and ultraviolet (UV) spectra 
of background quasars and are presumably the high-$z$ counterparts of today's normal galaxies. 
Further, while the \hi\ 
content of high-$z$ DLAs is insufficient (by about a factor of 2) to entirely fuel the stars 
observed in today's galaxies \citep[e.g.][]{prochaska05,noterdaeme09}, DLAs contain the bulk of 
the neutral hydrogen at high redshifts \citep[$\approx 80$\%; e.g.][]{omeara07}.

The fact that DLAs are selected to lie along the sightline to bright quasars implies that it is 
difficult to directly image the host galaxies, and estimate their stellar masses and star 
formation rates (SFRs). Indeed, despite numerous searches, galaxy counterparts have been identified 
and/or SFRs estimated for only about ten DLAs at $z \gtrsim 2$ 
\citep[see, e.g., ][and references therein]{krogager12}, 
with a few more systems identified at $z \approx 1$ \citep[e.g.][]{peroux12}. It is also clear from
these studies that typical high-$z$ DLAs have low SFRs, $\lesssim 1-10 M_\odot$~yr$^{-1}$, making
them additionally difficult to detect in optical/near-infrared line or continuum emission \citep[e.g.][]{fumagalli10}. 
Finally, sensitivity issues with current radio telescopes have precluded the detection of, or even 
useful limits on, radio \hii\ or CO line emission from high-$z$ DLAs \citep[e.g.][]{wiklind94,kanekar06}.

Most of our information on DLAs at all redshifts hence stems from absorption spectroscopy. For example,
optical and UV spectroscopy have yielded estimates of metallicities and abundances in a 
large number of DLAs out to $z \sim 5$
\citep[see][and references therein]{rafelski12}. 
DLA metallicities are now known to evolve with redshift, with the average metallicity higher at 
low redshifts \citep[e.g.][]{prochaska03a,rafelski12}; however, low metallicities are typical
at all redshifts. The correlations observed between DLA metallicities and both the velocity widths or 
rest equivalent widths of low-ionization metal lines \citep[][]{wolfe98,ledoux06,prochaska08} suggest 
the presence of a mass-metallicity relation in DLAs \citep[e.g.][]{moller13,neeleman13}, 
as has been observed in other high-$z$ galaxies 
\citep[e.g.][]{tremonti04}. High-$z$ DLAs have been shown to have low molecular fractions, typically 
$\lesssim 10^{-5}$, with only one-sixth showing detectable ${\rm H_2}$ absorption 
\citep[e.g.][]{ledoux03,noterdaeme08}. High-$z$ DLAs also have low fractions of cold atomic gas, 
with most of the \hi\ in the warm phase for absorbers at $z \gtrsim 2$ 
\citep[e.g.][]{kanekar03,kanekar14}, apparently due to the low DLA metallicities and hence, 
the paucity of cooling routes \citep[][]{kanekar01a,kanekar09c}. Low-ionization metal
lines detected in DLAs tend to have multiple components, sometimes with very different relative 
abundances \citep[e.g.][]{dessauges06,kanekar06}, and large velocity widths, $\gtrsim 90$~\kms\
\citep{prochaska97}.

While the above results provide insight on physical conditions in DLAs, they only trace conditions
along the narrow pencil beam towards the background QSO. There is hence often ambiguity in their
interpretation, even in terms of the big-picture view \citep[e.g. large disk galaxies versus 
small merging systems; ][]{prochaska97,haehnelt98}. As a result, we as yet have very little 
information on the basic physical characteristics of high-$z$ DLAs, especially their mass and size.

At low redshifts, Hubble Space Telescope and ground-based imaging of DLA fields have found DLAs to 
arise in a range of galaxy types, from dwarfs to large spiral disks \citep[e.g.][]{lebrun97,rao03,chen05}, 
consistent with the general galaxy population \citep[e.g.][]{zwaan05}. However, these results have 
been limited both by the small sample size and by concerns that a fainter galaxy at yet smaller 
impact parameter (i.e. beneath the quasar) might be responsible for the DLA absorption. As such, 
the crucial missing pieces in our understanding of DLA evolution remain the masses and sizes of 
the absorbing galaxies. 

Low-$z$ DLAs offer the exciting possibility of resolving the above issues through a direct 
measurement of the \hi\ mass, the spatial extent of the neutral gas and the gas velocity field, 
via \hii\ emission observations. Unlike optical imaging studies, which, even for low-$z$ DLAs are 
typically stymied by the quasar \citep[although see][]{fumagalli10}, \hii\ observations permit 
detections of galaxies at all impact parameters from the quasar sightline. And, of course, it would 
be possible to relate the measured properties to the characteristics of the DLA obtained from 
optical/UV absorption spectroscopy (e.g. metallicities, abundances, etc) and optical/UV imaging 
(e.g. star formation rates, stellar masses, etc) to obtain a comprehensive picture of the absorbing 
galaxies. 

Until recently, the weakness of the \hii\ transition as well as the lack of damped and sub-damped 
Lyman-$\alpha$ absorbers at low redshifts, $z \lesssim 0.2$ had limited searches for \hii\ emission to 
three absorbers, at $z = 0.0063$ towards PG~1216+069 \citep{kanekar05c,briggs06},  at $z = 0.009$ 
towards SBS1549+543 
\citep[][]{bowen01b,chengalur02} and $z = 0.101$ towards PKS~0439$-$433 \citep[][]{kanekar01e}. The 
high far-ultraviolet sensitivity of the new {\it Cosmic Origins Spectrograph} ({\it COS}) onboard the 
{\it Hubble Space Telescope} ({\it HST}) has resulted in the detection of three new DLAs and sub-DLAs
at $z \approx 0.1$ \citep{meiring11}. In this {\it Letter}, we report a search 
for redshifted \hii\ emission from these absorbers, at $z = 0.0830$ towards J1553+3548, $z = 0.0963$ 
towards J1619+3342 and $z = 0.1140$ towards J1009+0713, using the Green Bank Telescope (GBT).

\section{Observations, data analysis and results}
\label{sec:obs}

\begin{figure*}
\begin{center}
\includegraphics[scale=0.4,angle=0]{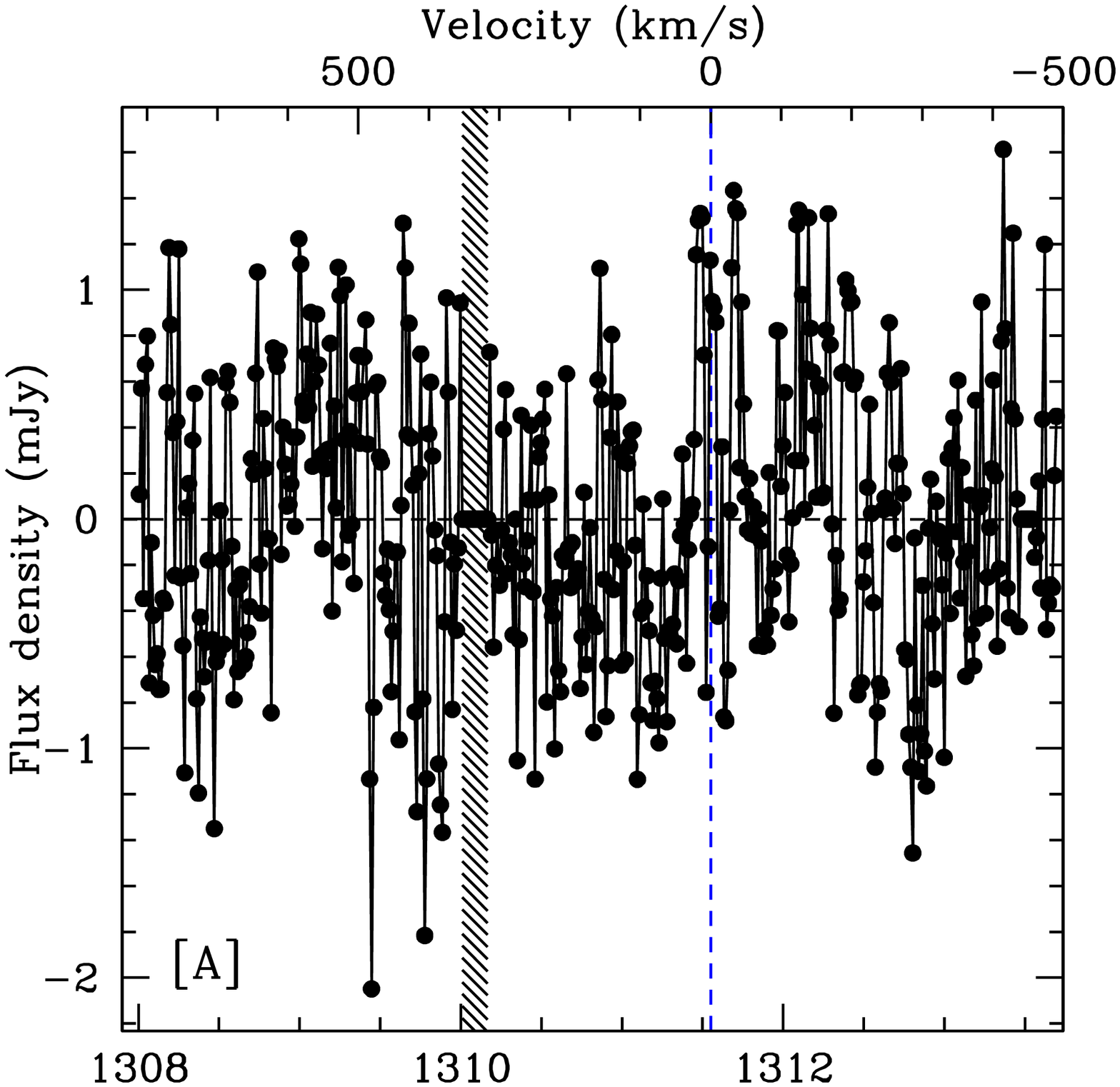}
\includegraphics[scale=0.4,angle=0]{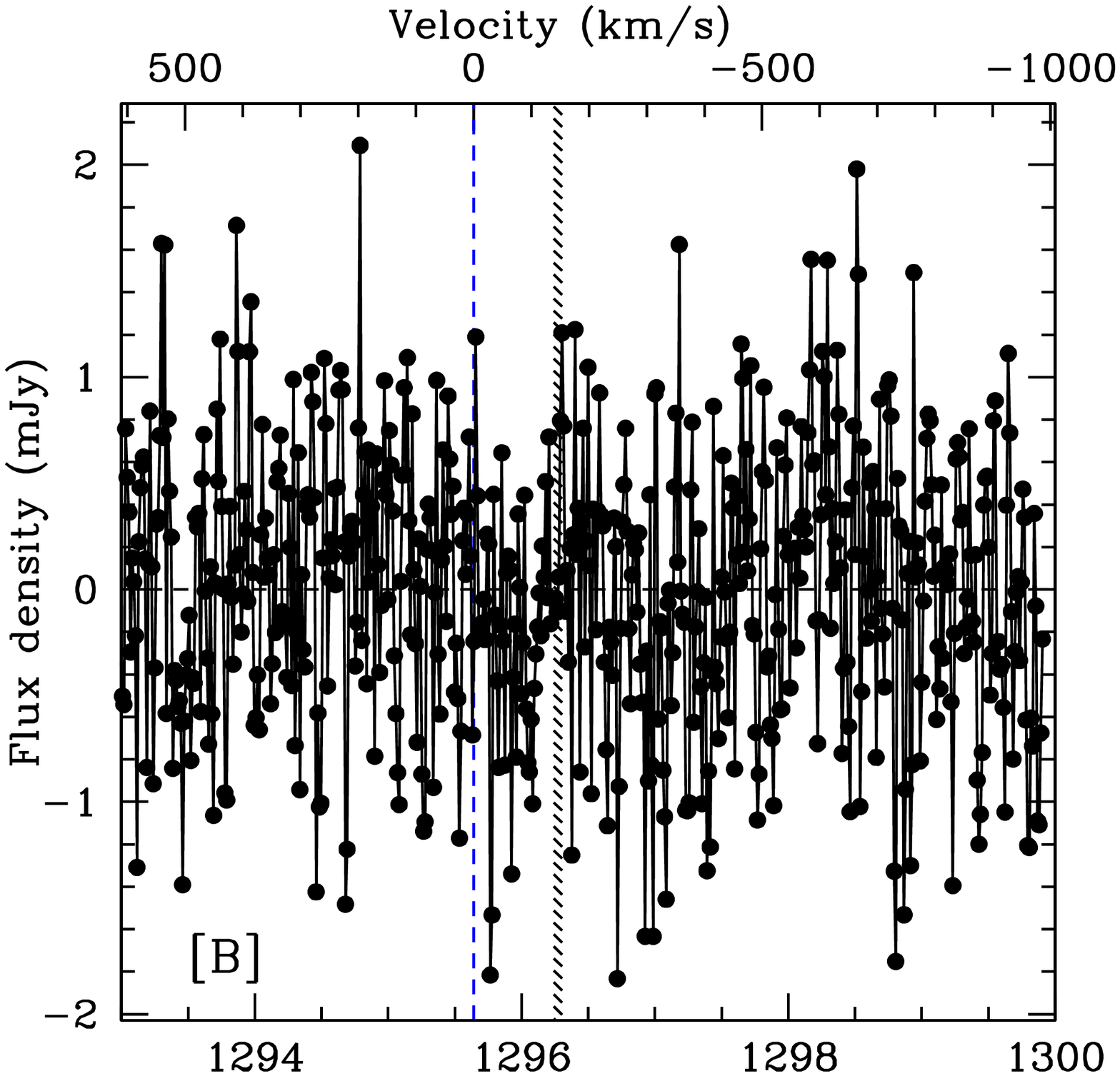}
\includegraphics[scale=0.4,angle=0]{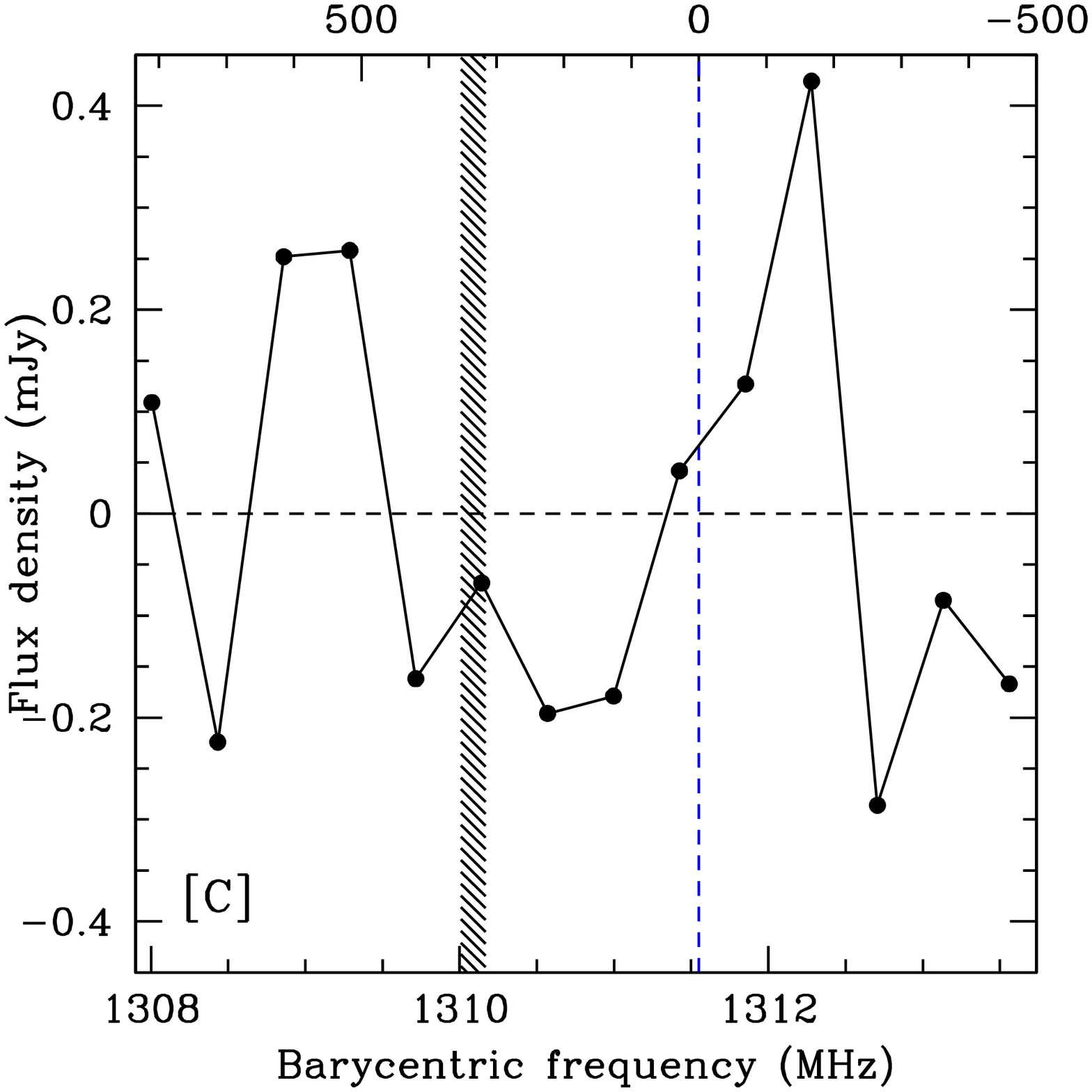}
\includegraphics[scale=0.4,angle=0]{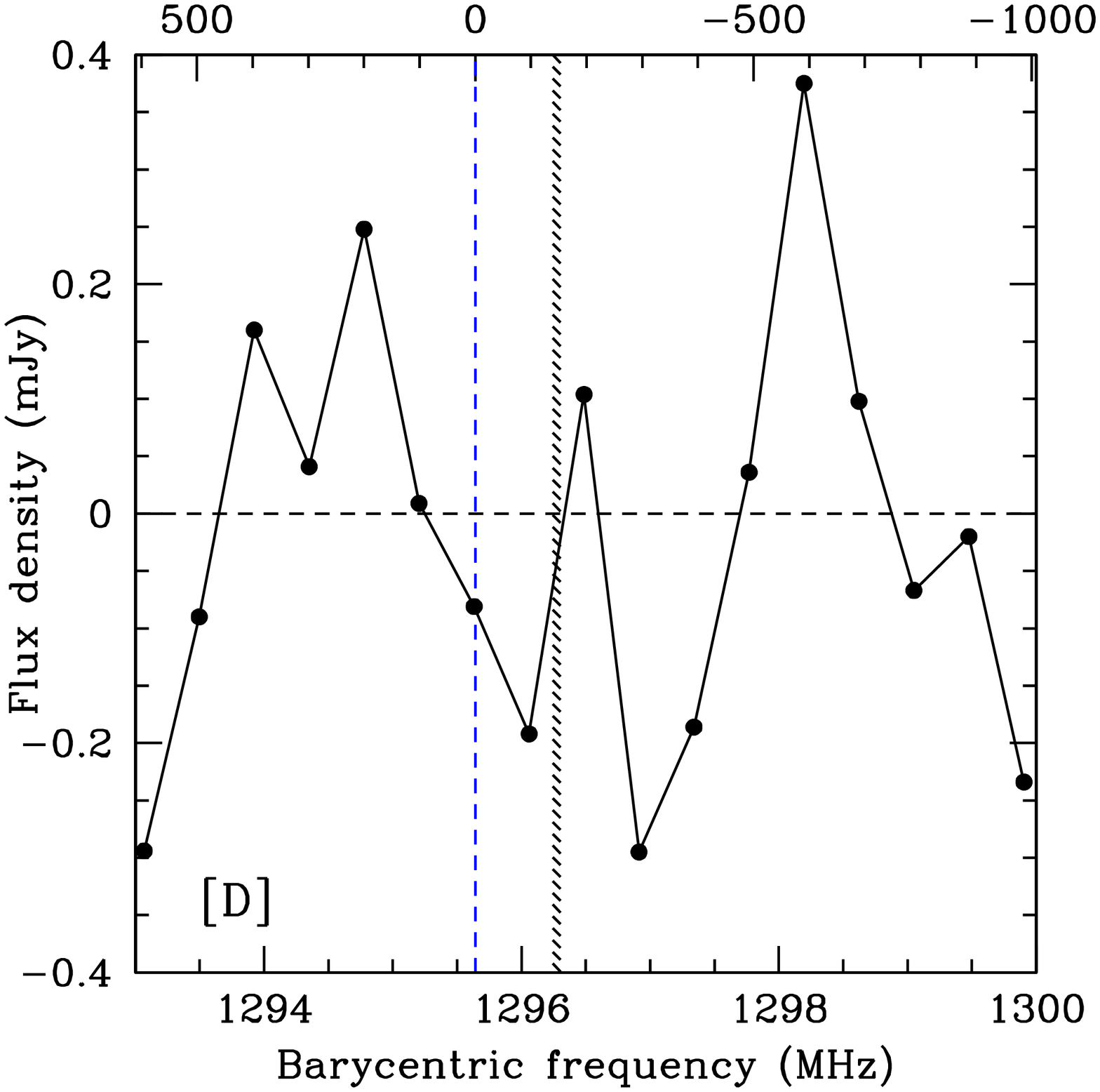}
\caption{GBT \hii\ spectra from the $z = 0.0830$ sub-DLA towards J1553+3548 (left panels) and the 
$z = 0.0963$ DLA towards J1619+3342 (right panels). The x-axis of each panel is barycentric frequency 
(in MHz), the y-axis is flux density (in mJy), while the top axis shows velocity (in \kms), relative to 
the absorber redshift. The top panels ([A] and [B]) show the spectra at a velocity resolution of 2.8~\kms, 
after Hanning-smoothing and re-sampling; the lower two panels ([C] and [D]) show the spectra after 
smoothing to a velocity resolution of $\approx 100$~\kms. The shaded regions indicate frequency ranges 
affected by RFI, while the dashed vertical line marks the expected location of the \hii\ emission line.}
\label{fig:spectra}
\end{center}
\end{figure*}

The GBT L-band receiver was used for the observations between June and September 2011, with the 
AutoCorrelation Spectrometer (ACS) as the backend, four 50\,MHz ACS sub-bands, two linear 
polarizations, and 10-second records. The same frequency settings were used for all four absorbers, 
with three ACS sub-bands centred on the three redshifted \hii\ line frequencies, so as to test for 
radio frequency interference (RFI). Each ACS sub-band was divided into 8192 spectral channels, 
yielding a velocity resolution of $\approx 1.4$~\kms\ and a total velocity coverage of 
$\approx 11,500$~\kms\ for each absorber. Position switching was used to calibrate the passband 
shape, with a 10-min. On/Off cycle, while the flux density scale was determined via online 
system temperature measurements with a blinking noise diode. The total observing time was 9~hours 
(J1553+3548), 11~hours (J1619+3342) and 33~hours (J1009+0713). Note that more time was spent on 
the $z = 0.1140$ DLA towards J1009+0713 as strong RFI was detected during the observations at 
$\approx 1275$\,MHz, close to its redshifted \hii\ line frequency.

All data were analysed using {\sc dish}, the {\sc aips++} single-dish package. Detailed data editing 
was necessary, due to the presence of significant amounts of RFI in all datasets as well as ACS 
failures. This was done via visual inspection of the calibrated dynamic spectra. About 20\% 
of the data had to be edited out for the absorbers towards J1553+3548 and J1619+3342, while the 
data towards J1009+0713 were entirely unusable. After the data editing, the individual spectra were 
calibrated, with a second-order baseline fit to each 10-second spectrum and subtracted out. The 
Hanning-smoothed and re-sampled spectra were then averaged together, to produce a spectrum for 
each observing epoch. For each source, the spectra from different epochs were shifted to the 
barycentric frame and then averaged together, using weights based on the root-mean-square (RMS) 
noise values. The final spectra for the absorbers towards J1553+3548 and J1619+3342 had RMS noise 
values of $\approx 0.65$~mJy per 2.8~\kms\ channel; these are shown in the two top panels of 
Fig.~\ref{fig:spectra}. 

The search for redshifted \hii\ emission was carried out after smoothing the spectra to, 
and re-sampling at, velocity resolutions of $\approx 20$, $40$, and $100$~\kms. The lower 
panels of Fig.~\ref{fig:spectra} show the two \hii\ spectra at velocity resolution of $\approx 100$~\kms.
The grey shaded regions indicate frequency ranges affected by narrow-band RFI, while the dashed vertical
line marks the expected redshifted \hii\ line frequency. It is clear that there is no evidence for 
either RFI or \hii\ line emission within $\approx 350$~\kms\ of the expected redshifted line frequencies.
The RMS noise values at a velocity resolution of $\approx 100$~\kms\ are 0.22~mJy (J1553+3548) and 
0.19~mJy (J1619+3342).

\section{Discussion}
\label{sec:res}

\setcounter{table}{0}
\begin{table*}
\begin{center}
\caption{\label{table:sample} Summary of searches for \hii\ emission in DLAs and sub-DLAs.}
\begin{tabular}{|c|c|c|c|c|c|c|}
\hline 
QSO 	    	& $z_{\rm DLA}$ & $\nhi$		& \mhi$^a$	      & [Z/H]$^b$        & Reference \\
	    	& 	 	& $\times 10^{20}$~\cm	& M$_\odot$~ 	      & 		 &  \\
\hline 
	    	& 	 	& 			&		      & 		 & \\
PG~1216+069   	& 0.0063   	& 0.2     		& $1.5 \times 10^7$   &  $-1.60 \pm 0.10$ & 1,2,11	  \\
SBS~1543+543    & 0.009         & 2.2 			& $1.3 \times 10^9$   & $-0.41 \pm 0.06$ & 3,4,5 \\
SDSS~J1553+3548 & 0.0830	& 0.35			& $< 2.3 \times 10^9$ & $-1.10 \pm 0.16$ & 6,7,8 \\ 
SDSS~J1619+3342 & 0.0963	& 3.5			& $< 2.7 \times 10^9$ & $-0.62 \pm 0.13$ & 6,7,8 \\
PKS~0439$-$433 	& 0.101		& 0.71			& $< 3.7 \times 10^9$ & $-0.42 \pm 0.12$ & 9,10 \\
	    	& 	 	& 	& 			& 	    \\
\hline 
\end{tabular}
\end{center}
\begin{center}
(1)~\citealp{kanekar05c}; (2)~\citealp{briggs06}; (3)~\citealp{bowen01b}; 
(4)~\citealp{chengalur02}; (5)~\citealp{bowen05}; (6)~\citealp{meiring11}; (7)~this work; 
(8)~\citealp{battisti12}; (9)~\citealp{kanekar01e}; (10)~\citealp{chen05}; (11)~\citealt{tripp05}. \\
$^a$~The \hi\ masses have been scaled to the $\Lambda$CDM cosmology used in this paper, 
with $3\sigma$ mass limits quoted at a velocity resolution of $100$~\kms. \\
$^b$~The quoted metallicities [Z/H] are based on [0/H]~\citep[PG1216+069; ][]{tripp05}, 
[S/H]~\citep[SBS1543+593 and SDSS~J1619+3342; ][]{bowen05}, 
[Si/H]~\citep[SDSS~J1553+3548; ][]{battisti12} and 
[Fe/H]+0.3~\citep[PKS~0439$-$433; ][]{chen05,rafelski12}.
\end{center}
\end{table*}

The three DLAs and sub-DLAs that were observed with the GBT were discovered via {\it HST-COS}
ultraviolet spectroscopy of a set of background quasars, selected for studies of gas at 
entirely different redshifts \citep{meiring11,tumlinson13}. The absorber redshifts are
well-constrained by both the Ly$\alpha$ and associated metal-line transitions to an accuracy 
better than a few tens of \kms\ \citep{battisti12}. The metal-line transitions also 
enable estimates of the gas-phase elemental abundances, and hence, the absorber metallicities.
The three absorbers studied here have metallicities (relative to solar) 
of [Z/H]~$= -1.12 - -0.54$, using S or Si abundances, and making appropriate
ionization corrections for the sub-DLA towards J1553+3548 \citep{battisti12}.

Our GBT non-detections of \hii\ line emission can be used to place constraints on the total 
\hi\ mass of the two absorbers, using the relation \citep[e.g.][]{rohlfs06}
\begin{equation}
{\rm M_{\rm HI}} = 2.356 \times 10^5 \times D_L^2 \times (1 + z)^{-1} \times \int S {\rm dV} \;\;,
\end{equation}
where \mhi\ (in M$_\odot$) is the \hi\ mass of the galaxy, $z$ is its redshift, $D_L$ (in Mpc) is 
its luminosity distance\footnote{We use a $\Lambda$-Cold Dark Matter ($\Lambda$CDM) cosmology, with 
H$_0 = 67.3$~\kms~Mpc$^{-1}$, $\Omega_m = 0.315$ and $\Omega_\Lambda = 0.685$ \citep{planck13}.}, 
and $\int S {\rm dV}$ (in Jy~\kms) is the velocity integral of the \hii\ line flux density.

For non-detections of \hii\ emission, one must assume a velocity distribution for the 
emitting gas to derive limits on the \hi\ mass. Note that, for a given \hi\ mass, a galaxy with 
a narrower \hii\ emission profile would yield a lower peak line flux density. We will 
assume the \hii\ emission profiles to be Gaussian, with a line full-width-at-half-maximum 
$\Delta V = 100$~\kms.  This yields $3\sigma$ upper limits of $\int S {\rm dV} < 
0.060 \times (\Delta V/100)^{1/2}$~Jy~\kms\ (J1553+3548) and $\int S {\rm dV} < 0.069 \times (\Delta V/100)^{1/2}$~Jy~\kms\ (J1619+3342) on the integrated \hii\ line flux density, and  
\mhi~$ < 2.3 \times 10^9 \times (\Delta V/100)^{1/2}$~M$_\odot$ (J1553+3548) and 
\mhi~$< 2.7 \times 10^9 \times (\Delta V/100)^{1/2}$~M$_\odot$ (J1619+3342) 
on the \hi\ masses of the two absorbers.

Table~\ref{table:sample} summarizes our current knowledge of the atomic gas masses of 
low-$z$ DLAs and sub-DLAs. The sample consists of five systems, two DLAs and three 
sub-DLAs. \hii\ emission has been detected in only two objects, the $z \approx 0.0063$ 
sub-DLA towards PG~1216+069 \citep{briggs06} and the $z \approx 0.009$ DLA towards 
SBS~1549+543 \citep[][]{bowen01b}. Table~\ref{table:sample} also lists the \hi\ 
column densities of the absorbers and (when available) their metallicities and 
velocity widths.

It is clear from Table~\ref{table:sample} that the inferred \hi\ masses or limits on the 
\hi\ mass are in all cases significantly lower (by a factor of at least a few) than the 
\hi\ mass at the knee of the Schechter function that provides a good fit to the \hi\ 
mass distribution of galaxies in the local Universe, log[M$^*_{\rm HI}/{\rm M}_\odot] = 
(9.96 \pm 0.02)$  \citep[][]{martin10}, using H$_0 = 67.3$~\kms~Mpc$^{-1}$. Thus, 
although the number of damped systems that have so far been searched for \hii\ 
emission is still quite small, the present data indicate that low-redshift DLAs and 
sub-DLAs do not typically arise in massive, gas-rich galaxies.

It is interesting to compare our results with the predictions of \citet{zwaan05} for  
the typical \hi\ mass of low-$z$ DLAs. \citet{zwaan05} used Westerbork Synthesis Radio 
Telescope \hii\ emission images of a large sample of nearby galaxies to test whether 
high-$z$ DLAs might arise in galaxies with gas distributions similar to that in the 
$z = 0$ galaxy population. They found both the incidence rate, and the distribution of 
impact parameters and \hi\ column densities, of low-$z$ DLAs to be consistent with 
a scenario in which the absorbers arise in the local galaxy population. The median 
$z=0$ DLA is expected to arise in an low-luminosity (L*/7) galaxy, with an \hi\ 
mass of $2 \times 10^9$~M$_\odot$. While high-mass DLAs do contribute significantly
to the cross-section, it is the highest \hi\ column densities that tend to arise in 
the most massive galaxies. Comparing with our results, it is clear from 
Table~\ref{table:sample} that the five DLAs and sub-DLAs at $z \lesssim 0.1$ that have 
so far been searched for \hii\ emission all have low $\nhi$ values, $\leq 3.5 \times 
10^{20}$~\cm, and low \hi\ masses, $< 3.7 \times 10^9$~M$_\odot$. Our results are thus 
entirely consistent with the predictions of \citet{zwaan05}. It would be interesting to 
search for \hii\ emission from low-$z$ DLAs with high \hi\ column densities, for 
which \citet{zwaan05} expect the cross-section to be dominated by massive galaxies.

There have been suggestions in the literature that sub-DLAs (also known as super-Lyman
limit systems, with $10^{19}$~\cm~$\leq \nhi \leq 2 \times 10^{20}$~\cm) arise in 
galaxies that are systematically more massive than the galaxies that give rise to DLAs 
\citep[e.g.][but see \citealp{dessauges09}]{khare07,kulkarni10}. We emphasize that the 
three sub-DLAs of Table~\ref{table:sample} also have low \hi\ masses, $< 3.7 \times 
10^9 M_\odot$, contradicting the above arguments. Of course, our conclusion is tempered by
sample variance and must be tested with a much larger survey.  

Finally, the metallicity of DLAs is known to show a correlation with the velocity width 
$V_{\rm 90}$ of unsaturated low-ionization metal lines \citep{wolfe98,ledoux06},
as well as with the rest equivalent width of saturated metal lines \citep{prochaska08}.
These correlations have usually been interpreted in terms of a mass-metallicity relation 
in the absorbers \citep[e.g.][]{ledoux06,moller13,neeleman13}, similar to the mass-metallicity 
relation that has been found in high-$z$ emission-selected galaxies 
\citep[e.g.][]{tremonti04,erb06,maiolino08}. Three of the absorbers of 
Table~\ref{table:sample} have high metallicities, [Z/H]~$\geq -0.62$, in the top 10\% of 
DLA metallicities at all redshifts \citep[e.g.][]{rafelski12}. It is interesting that 
all three of these high-metallicity absorbers show low \hi\ masses, 
M$_{\rm HI} < 3.7 \times 10^9$~M$_\odot$. We thus find that absorbers with high 
metallicities do not necessarily have high gas masses, in apparent contradiction 
with the putative mass-metallicity relation. However, it should be emphasized that the 
metallicity estimates only probe a pencil beam through the three galaxies; testing 
the mass-metallicity relation would require a larger sample. Further, the mass-metallicity 
relation for emission-selected galaxies involves the {\it stellar} mass while that 
for DLAs involves the {\it virial} mass (dominated by dark matter), neither of 
which can be immediately linked to the gas mass. 

In summary, we have carried out a deep GBT search for redshifted \hii\ emission from 
three DLAs and sub-DLAs at $z \approx 0.1$. Our GBT non-detections of \hii\ emission yield
tight constraints on the \hi\ mass of two absorbers, \mhi~$< 2.3 \times 10^9 \times 
(\Delta V/100)^{1/2}$~M$_\odot$ for the $z = 0.0830$ sub-DLA towards J1553+3548, and 
\mhi~$< 2.7 \times 10^9 \times (\Delta V/100)^{1/2}$~M$_\odot$ for the $z = 0.0963$ 
DLA towards J1619+3342. The data on the third absorber, at $z = 0.114$ towards J1009+0713, 
was rendered unusable by strong RFI close to the redshifted \hii\ line frequency. Our 
results are consistent with a scenario in which most low-\hi\ column density DLAs and 
sub-DLAs at low redshifts do not arise in massive gas-rich galaxies.

\section*{Acknowledgements}

PM acknowledges support from the NCRA-TIFR Visiting Students' Research Programme, 
and NK from the Department of Science and Technology through a Ramanujan Fellowship. 
JXP is partly supported by NSF grant AST-1109447. The NRAO is a facility of the 
National Science Foundation operated under cooperative agreement by Associated 
Universities, Inc.. We thank an anonymous referee for a careful reading of an 
earlier version of this paper.

\label{lastpage}

\bibliographystyle{mn2e}
\bibliography{ms}

\end{document}